\documentclass[twocolumn,showpacs,preprintnumbers,amssymb,pra]{revtex4}

\usepackage{amssymb}
\usepackage{amsmath}
\usepackage{graphicx}
\usepackage{dcolumn}
\usepackage{bm}
\usepackage{verbatim}
\usepackage{array}
\usepackage{epsfig}
\usepackage{amsbsy}
\usepackage{color}
\usepackage{soul}
\usepackage{url}
\usepackage{hyperref}
\usepackage{longtable}

\usepackage[mediumqspace,mediumspace,amssymb,Gray]{SIunits}

\newcommand{\ds}{\displaystyle}
\newcommand{\ddsum}[1]{{\displaystyle \sum_{ #1 }}}

\def\bra#1{\mathinner{\langle{#1}|}}
\def\ket#1{\mathinner{|{#1}\rangle}}

\begin{document}
\preprint{\hfill\parbox[b]{0.3\hsize}{ }}

\title{Quantum interference in laser spectroscopy of highly charged lithiumlike ions}
 
\author{ 
        Pedro Amaro$^{1} \footnote{pdamaro@fct.unl.pt}$,
        Ulisses Loureiro$^{1}$,
        Laleh Safari$^{2}$,
        Filippo Fratini$^{3}$, \\
        Paul~Indelicato$^{4}$,
        Thomas~St\"{o}hlker$^{5,6, 7}$,
        Jos\'e Paulo Santos$^{1}$
          }
\affiliation
{\it
$^{1}$ Laborat\'orio de Instrumenta\c{c}\~ao, Engenharia Biom\'edica e F\'isica da Radia\c{c}\~ao
(LIBPhys-UNL),~Departamento de F\'isica, Faculdade~de~Ci\^{e}ncias~e~Tecnologia,~FCT,~Universidade Nova de Lisboa,~2829-516 Caparica, Portugal \\
$^{2}$ IST Austria, Am Campus 1, A-3400 Klosterneuburg, Austria \\
$^{3}$ University of Applied Sciences BFI Vienna, Wohlmutstra{\ss}e 22, A-1020 Vienna \\
$^{4}$ Laboratoire Kastler Brossel, Sorbonne Universit\'e, CNRS, ENS-PSL Research University, Coll\`ege de France, Case\ 74;\ 4, place Jussieu, F-75005 Paris, France \\
$^{5}$ GSI Helmholtzzentrum f\"{u}r Schwerionenforschung, D-64291 Darmstadt, Germany\\
$^{6}$  Helmholtz-Institut Jena, D-07743 Jena, Germany\\ 
$^{7}$ Institut f\"{u}r Optik und Quantenelektronik, Friedrich-Schiller-Universit\"{a}t Jena, Max-Wien-Platz 1, 07743 Jena, Germany
}

\date{\today}
\date{Received: \today  }


\begin{abstract}

We investigate the quantum interference induced shifts between energetically close states in highly charged ions, with the energy structure being observed by laser spectroscopy. In this work, we focus on hyperfine states of lithiumlike heavy-$Z$ isotopes  and quantify how much quantum interference changes the observed transition frequencies.  
  The  process of photon excitation and subsequent photon decay for the transition $2s\rightarrow2p\rightarrow2s$ is implemented with fully relativistic and full-multipole frameworks, which are relevant for such relativistic atomic systems. We consider the isotopes   $^{207}$Pb$^{79+}$ and $^{209}$Bi$^{80+}$ due to experimental interest, as well as other examples of isotopes with lower $Z$, namely $^{141}$Pr$^{56+}$ and $^{165}$Ho$^{64+}$. We conclude that  quantum interference can induce shifts up to 11\% of the linewidth in the measurable resonances of the considered isotopes, if interference between resonances is neglected. The inclusion of relativity decreases the cross section by  35\%, mainly due to the complete retardation form of the electric dipole multipole. However, the contribution of the next higher multipoles (e.g. magnetic quadrupole) to the cross section is negligible. This makes the contribution of relativity and higher-order multipoles  to the  quantum interference induced shifts a minor effect, even for heavy-$Z$ elements.  

\end{abstract}



\maketitle


\section{Introduction}
\label{intro}

In the near future , the Facility for Antiproton and Ion Research FAIR (located in Darmstadt, Germany) will allow lithium-like ions to be accelerated to relativistic speeds (reduced velocity of $\beta\approx0.9964$) at the heavy-ion ring-accelerator SIS100 \cite{bac2007, sga2014, sbb2015}. Atomic excitations of HCI  that are within the UV and soft x-ray range in the mass rest frame, such as the $2s_{1/2}\rightarrow2p_{1/2, 3/2}$,  can thus be measured by optical lasers in a collinear geometry, due to the relativistic Doppler effect. Additionally,  intense soft x-ray light sources are currently being developed for experiments at ion storage ring, which may even allow to address the $2s_{1/2}\rightarrow2p_{3/2}$ transition in Li-like uranium via laser spectroscopy \cite{rhd2015}. Such measurements of $2s_{1/2}\rightarrow2p_{1/2, 3/2}$ transitions can give further insight  about the recent and significant discrepancy (7$\sigma$) observed in the  ground-state hyperfine splitting of H-like and  Li-like bismuth \cite{jzc2017}.
%

One systematic effect in laser spectroscopy, which has been recently investigated in neutral atoms, is the induced shifts by overlapping resonances, often referred as quantum interference (QI). As pointed in the seminal work of Low \cite{low1952}, this QI between the main resonant channel and other non-resonant channels leads to an asymmetry of the line profile, which restricts the use of standard Lorentz line profile for representing the line profile with accuracy. Otherwise, these asymmetries induces shifts that depend on the relative distance between neighborhood resonances and respective linewidths. Although these shifts were often negligible \cite{low1952}, there has been some attention paid by the high-precision spectroscopy community of neutral atomic systems. With the current level of accuracy achieved  in such high-precision  experiments, a careful analysis of QI contribution is mandatory. Therefore, several calculations of QI shifts have  been made for current high-precision laser and microwave spectroscopy of fine and hyperfine structure in many atomic systems. Investigations of QI shifts performed in hydrogen transition $1s-2s$ \cite{lsp2001, jem2002, lss2007, hbl2017},  cesium \cite{tam2015}, ytterbium \cite{kgb2016} and muonic atoms hyperfine structures \cite{abj2015, afs2015} found a negligible QI contribution with the respective experimental accuracy.  However, it is important to notice that investigations with hydrogen transition $1s-3s$ \cite{mpb2014, bmk2015, hbl2017}, the helium fine structure \cite{hoh2010, mhh2012,mhh2012b, mhh2015, fzs2015, mhh2017} and lithium hyperfine structure \cite{scs2011,   bup2013}  found that QI contribution was the cause of discrepancy between data and predictions. Moreover, recent investigations \cite{bup2013, afs2015} have shown a clear dependency between the QI shift and the angular and polarization conditions of measurement, which can be employed for minimization, and even removal, of QI shifts in laser spectroscopy. 

 \begin{figure}[t]
  \centering
\includegraphics[clip=true,width=1.0\columnwidth]{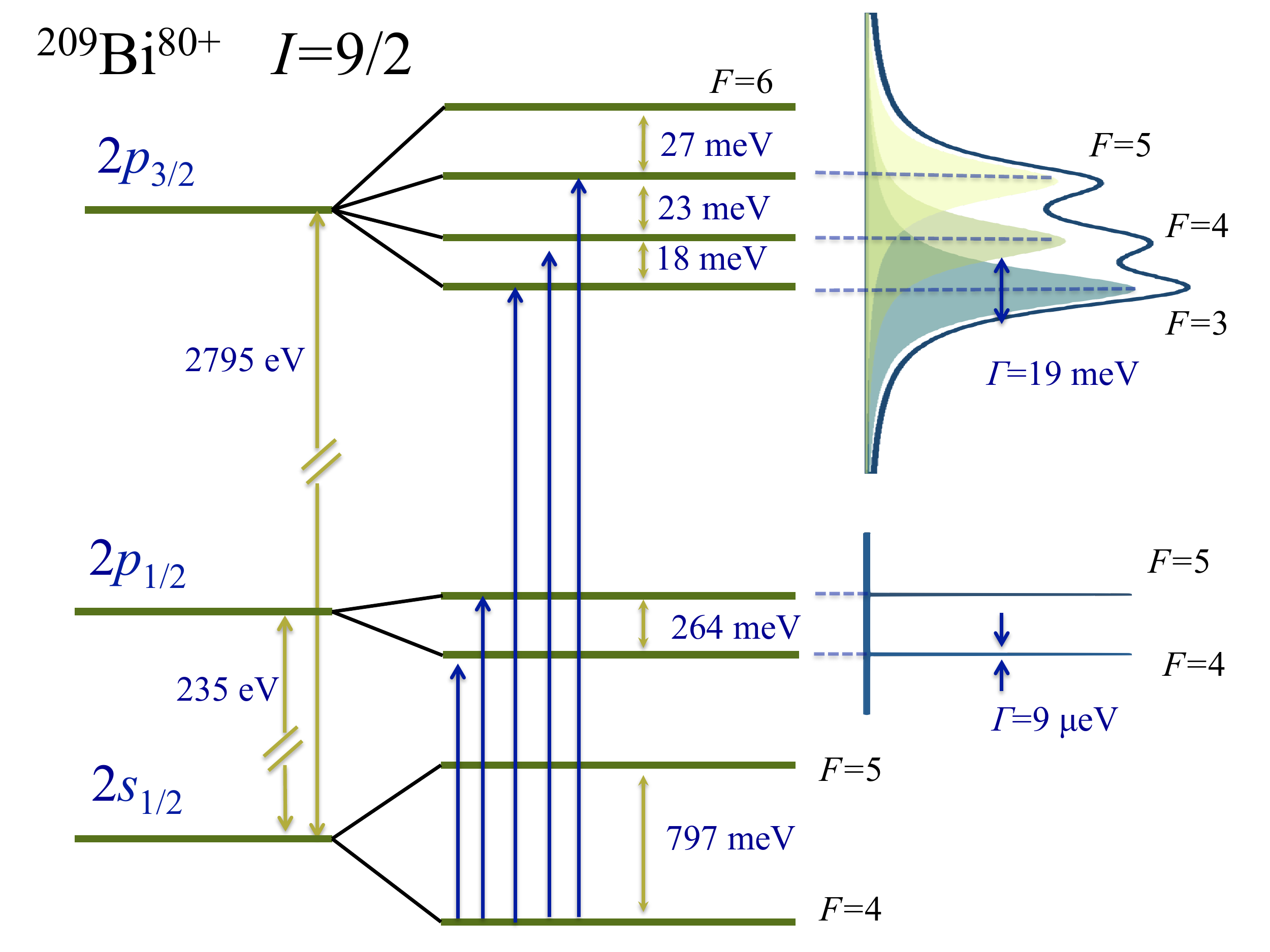}
\caption{ (Color online) $2s$ and $2p$ hyperfine structure for $^{209}$Bi$^{80+}$.  Allowed electric dipole resonances are shown for initial $2s_{1/2}^{F_i=4}$ states. 
} 
\label{fig:transit}
\end{figure}

Hyperfine splittings in isotopes of lithiumlike HCIs  ranges from few tens to hundreds of meV \cite{bes1999, kos2005} for both $2p_{1/2}^F$ and $2p_{3/2}^F$ states along all the isoelectronic $Z$ sequence ($F$ is the total angular momentum). Since the linewidths $\Gamma$ of $2p_{1/2}^F$ are always lower than tens of \micro eV \cite{tcl1991} for all lithiumlike HCIs,  it is expected that QI plays a small role. According to the rule-of-thumb for distant resonances (QI$\sim \Gamma^2   / 4 \Delta$ with $\Delta$ being the energy difference between two resonances) \cite{hoh2010}, typical QI shifts for $2p_{1/2}^F$ states are roughly 0.02\% of $\Gamma$. On the other hand, linewidths of $2p_{3/2}^F$ states increase according to $Z^4$ along the isoelectronic sequence, which can reach  tens of meV for heavier ions, such as $^{207}$Pb$^{+79}$ and $^{209}$Bi$^{+80}$ (cp. Fig.~\ref{fig:transit}). A careful analysis of the QI shifts is thus required. 

\begin{table}[t]
\caption{\label{tab:parame_hyper_1} Values of excitation energies of  $2p_{3/2}$ ($\Delta E_{2p_{3/2}}$)  \cite{tcl1991}, magnetic-dipole hyperfine splittings $\mathcal{A}$ \cite{bes1999, kos2005},  as well as the linewidth $\Gamma$ \cite{tcl1991} for  $^{141}$Pr$^{56+}$, $^{165}$Ho$^{64+}$, $^{207}$Pb$^{79+}$ and  $^{209}$Bi$^{80+}$. Values in meV are given in the rest frame. 
} 
\begin{ruledtabular}
\begin{tabular}{lrrrr}
   & \multicolumn{1}{c}{$^{141}$Pr$^{56+}$}  & \multicolumn{1}{c}{$^{165}$Ho$^{64+}$} & \multicolumn{1}{c}{$^{207}$Pb$^{79+}$}& \multicolumn{1}{c}{$^{209}$Bi$^{80+}$} \\
   \\[-2.0ex]  \cline{2-3} \cline{4-5}  \\[-2.0ex]	\\[-2.0ex]	
 $\Delta E_{2p_{3/2}}$ 		   	    	                          	&	6.841$\times10^{5\phantom{-}}$	&	1.124$\times10^{6\phantom{-}}$	&	 2.649$\times10^{6\phantom{-}}$ 	&	 2.795$\times10^{6\phantom{-}}$ 	\\
 $\mathcal{A}_{2s_{1/2}}$       	                                   	&	8.260$\times10^{1\phantom{-}}$	&	1.350$\times10^{2\phantom{-}}$	&	 7.858$\times10^{1\phantom{-}}$ 	&	 3.589$\times10^{2}$\footnote{Obtained from the measured hyperfine splitting in  Ref.~\cite{jzc2017}.}\hspace{0.05cm}  	\\
  $\mathcal{A}_{2p_{3/2}}$                                                   	&	10.68$\times10^{1\phantom{-}}$	&	15.73$\times10^{1\phantom{-}}$	&	 4.308$\times10^{0\phantom{-}}$ 	&	 3.107$\times10^{1\phantom{-}}$ 	\\
  $\Gamma_{2p_{3/2}}$                 	&	5.825$\times10^{-1}$	&	1.936$\times10^{0\phantom{-}}$	&	 1.605$\times10^{1\phantom{-}}$ 	&	 1.881$\times10^{1\phantom{-}}$ 	\\
\end{tabular}
\end{ruledtabular}
\end{table}

The aim of this investigation is to extend the treatment of the QI shift,  recently performed in muonic systems \cite{abj2015, afs2015}, to lithiumlike HCIs. To the best of our knowledge, a relativistic and full-multipole framework is  implemented for the first time. 
In view of the possibility of laser spectroscopy of the $2s_{1/2} \rightarrow2p_{3/2}$ transition lithiumlike HCI \cite{bac2007}, we calculate  the respective QI shifts transitions in lithiumlike ions with hyperfine structure for the selected isotopes of 
  $^{141}$Pr$^{56+}$, $^{185}$Re$^{+72}$, $^{207}$Pb$^{79+}$ and $^{209}$Bi$^{80+}$.  The respective magnetic dipole splittings $\mathcal{A}$ in rest frame of the hyperfine states that is considered in this work was taken from QED calculations performed in Refs.~\cite{bes1999, kos2005} and are given in Tables \ref{tab:parame_hyper_1} and shown for $^{209}$Bi$^{80+}$ in Fig.~\ref{fig:transit}.  The  linewidths and excitation energies listed in Table~\ref{tab:parame_hyper_1} were taken from Ref.~\cite{tcl1991}. The theoretical formalism used here for evaluation of the QI shifts can be traced back to recent works published in Refs.~\cite{saf2012b, bup2013, abj2015, afs2015}.


Atomic units  are used throughout the article unless stated otherwise.

\section{Theory}
\label{sec:theor}

Laser spectroscopy is often modeled by the physical process of resonant photon scattering \cite{bup2013, bmk2015, abj2015, afs2015}. Here we consider the excitation of the $2s\rightarrow 2p$ transition by linearly polarized radiation and the detection of the $2p\rightarrow 2s$ fluorescence decay by polarization insensitive detectors. This process is often  described in the relativistic framework of the S-matrix theory \cite{alb1965, saf2012b} by the following  second-order general amplitude (velocity gauge) \cite{alb1965, saf2012b},
\begin{eqnarray}
 \ds \mathcal{M}_{i \rightarrow f} ^{ {\bm k_1}  {\bm \varepsilon_1}, {\bm k_2}   {\bm\varepsilon_2}  }(m_i, m_f ) =  \hspace{4cm}\nonumber \\
 \ds \ddsum{\nu}
\left[
\frac{ \bra{f} \hat{\mathcal{R}}^{\dag}(\bm k_2, \bm \varepsilon_2 )
\ket{\nu}\bra{\nu} \hat{\mathcal{R}}(\bm k_1,{\bm\varepsilon_1} ) \ket{i}}{\omega_{\nu i}
-\omega_1 -i \Gamma_{\nu} /2 }  \right] ~,
%
\label{Mfi}
\end{eqnarray}
Here, $\ket{i}$, $\ket{\nu}$ and $\ket{f}$ are the initial, intermediate and final hyperfine states of  the lithiumlike ion with a closed $1s$ shell.  Each state  has a well-defined total angular momentum, $F$, projection along the $z$ axis, $m$, and total atomic angular momentum $J$ of the uncoupled electron in the $n=2$ shell that participates in the transitions. $\omega_{\nu i} = E_\nu - E_i $ is the transition energy, or resonance, between  the intermediate $\ket{\nu}$ and initial $\ket{i}$ states. We consider here only transitions of the active electron, which reduces the calculation of the overall 3-electron wavefunction to a single hydrogenic case.  The region of incident photon energies studied in this work comprises the near-resonant region of the $2s-2p$ excitation, which validates  the restriction of the summation in Eq.~\eqref{Mfi} to only $\nu\equiv 2p_{J_\nu}^{F_\nu}$ intermediate states. 
Furthermore, the transition operator $\hat{\mathcal{R}}$ in Eq.~\eqref{Mfi} describes relativistic interaction between the active electron and the incident (scattered) photon with propagation vector ${\bm k_\gamma}$, polarization ${\bm \varepsilon_\gamma}$  and energy $\omega_\gamma$, which can be written as 
\begin{equation}
\hat{\mathcal{R}}(\bm{k}_{\gamma},\bm{\bm{\epsilon}_{\gamma}})=\bm{\alpha}\cdot\bm{\varepsilon_{\gamma}}e^{i\bm{k}_{\gamma}\cdot\bm{r}} \hspace{0.5cm} (\gamma=1,2)~.
\label{eq:R_opera}
 \end{equation}
Here, the $\bm{\alpha}$ and $\bm{r}$ denote the Dirac vector matrices and the vector position for the electron. The adjoint superscript in Eq.~\eqref{Mfi} for $\hat{\mathcal{R}}^{\dag}$  specifies the case of photon decay, while its absence addresses to a photon excitation case.

As in previous works \cite{abj2015, afs2015}, we adopted the experimental setting of a linearly polarized incident photon with nonobservation of the  scattered photonic polarization. With this setting, a polarization angle $\chi$ of the incident photon and a scattering angle $\theta$  are sufficient for describing the angular distribution of the scattering process \cite{abj2015}. The corresponding differential cross section of the amplitude in Eq.~\eqref{Mfi} with this experimental setting is given by \cite{alb1965}
\begin{eqnarray}
\frac{d\sigma}{d\Omega}(\omega, \theta, \chi) &=&  
 \frac{1}{(2F_{i}+1)} \times \nonumber \\   
&& \sum_{\tiny \begin{array}{c} m_{i},F_f \\ m_{f}, J_f \\ \bm{\varepsilon_{2}} \end{array} }\left|\mathcal{M}_{i \rightarrow f} ^{ {\bm k_1}  {\bm \varepsilon_1}, {\bm k_2}   {\bm\varepsilon_2}  }(m_i, m_f ) \right|^{2}~, 
 \label{eq:Msimpli}
\end{eqnarray} 
where $\omega \equiv \omega_1$  correspond to the energy of both incident and fluorescence photons.

Further simplification of the second-order amplitude \eqref{Mfi} can be achieved by considering a spherical tensor decomposition, also known as a multipole expansion, of the transition operator \eqref{eq:R_opera}.  For the absorption of a photon with direction $\bm{\hat{k}}_{\gamma}$, such a decomposition reads \cite{alb1965, god1981},
\begin{equation}
\bm{\alpha}\cdot \bm{\varepsilon}_{\gamma}e^{i\bm{k_{\gamma}}\cdot\bm{r}}=\sum_{LM\lambda}\left[\bm{\varepsilon}_{\gamma}\bm{\cdot}\bm{Y}_{LM}^{*(\lambda)}(\bm{\hat{k}}_\gamma)\right]\left[a_{LM}^{(\lambda)}(\bm{\hat{r}})\right]~,
\label{eq:multexpanc}
\end{equation}
where for electric ($\lambda=1$) and magnetic ($\lambda=0$) contributions, the $\bm{Y}_{LM}^{(\lambda)}(\bm{\hat{k}}_\gamma)$ are related to the vector spherical harmonic according to $\bm{Y}_{LM}^{(0)}(\bm{\hat{k}}_\gamma)=\bm{Y}_{LLM}(\bm{\hat{k}}_\gamma)$ and $\bm{Y}_{LM}^{(1)}(\bm{\hat{k}}_\gamma)= -i \bm{\hat{k}\times} \bm{Y}_{LM}^{(0)}(\bm{\hat{k}}_\gamma)$, respectively \cite{god1981}. Each multipole $a_{LM}^{(\lambda)}(\bm{\hat{r}})$ has an angular momentum $L$, angular momentum projection $M$, and parity $(-1)^{L+1+\lambda}$ and explicit forms of these terms can be found in Ref.~\cite{ god1981}. The first multipole of this expansion with $\lambda=1$ and $L=1$ is the often dominant electric dipole multipole ($E1$). The matrix elements of these multipole components $\langle f |\bm{ \alpha}\cdot a_{LM}^{(\lambda)*}(\bm{\hat{r}})| i \rangle$ are given in the Appendix. 

\begin{widetext}
For the purpose of investigating QI effects, the differential cross section \eqref{eq:Msimpli} can be rearranged in terms of Lorentezian terms $ \Lambda_{J_{i}J_{\nu}}^{F_{i} F_{\nu}}$ associated with the incoherent  part of the summation, as well as cross terms $\Xi_{J_{i}J_{\nu} J_{\nu'}}^{F_{i} F_{\nu} F_{\nu'}}$ that contains  the remained interference terms for a complete coherent summation. The result of this procedure (details can be found in Appendix) is given by 
\begin{equation}
\frac{d\sigma}{d\Omega}(\omega, \theta, \chi) = 
\frac{ \alpha^4 \omega^4 \mathcal{S}^4_{1s 2p_{3/2}}}{(2F_{i}+1)} 
 \left(
 \sum_{\small F_\nu,  J_\nu}  \frac{  \Lambda_{J_{i}J_{\nu}}^{F_{i} F_{\nu}}( \theta, \chi,\omega) }{(\omega_{\nu i} - \omega)^2 + (\Gamma_{\nu}/2 )^2 } +   \sum_{(F'_\nu,  J'_\nu) > (F_\nu,  J_\nu)}   \frac{\Xi_{J_{i}J_{\nu} J_{\nu'}}^{F_{i} F_{\nu} F_{\nu'}}( \theta, \chi,\omega)}{    (\omega_{\nu i} - \omega) (\omega_{\nu'  i} - \omega ) +  \Gamma_{\nu}\Gamma_{\nu'}/4}
 \right) ~.
 %
 \label{eq:Msimpli_2}
\end{equation}
 The obtained expressions of the Lorentzian terms $\Lambda_{J_{i}J_{\nu}}^{F_{i} F_{\nu}}( \theta, \chi,\omega)$, and cross terms $\Xi_{J_{i}J_{\nu} J_{\nu'}}^{F_{i} F_{\nu}  F_{\nu'}}( \theta, \chi,\omega)$ are given by 
\begin{eqnarray}
\Lambda_{J_{i}J_{\nu}}^{F_{i} F_{\nu}}( \theta, \chi, \omega)&=&   \sum_{m_{i},F_f,m_{f}, J_f, {\bm \varepsilon_{2} } }\left| \Omega_{J_{i}J_{\nu}J_{f}}^{F_{i} F_{\nu}F_{f}}( \theta, \chi, \omega, {\bm \varepsilon_{2} } ) \right|^2~,\label{eq:delta}\\
\Xi_{J_{i}J_{\nu} J_{\nu'}}^{F_{i} F_{\nu}  F_{\nu'}}( \theta, \chi, \omega)&=& 
2 \mbox{Re} \left[  \sum_{ \tiny \begin{array}{c} m_{i },F_f\\ m_{f} ,J_f, {\bm \varepsilon_{2} } \end{array}}  
\Omega_{J_{i}J_{\nu}J_{f}}^{F_{i}F_{\nu}F_{f}}( \theta, \chi, \omega, {\bm \varepsilon_{2} } ) 
 \Omega_{J_{i}J'_{\nu}J_{f}}^{F_{i} F'_{\nu}F_{f}}( \theta, \chi, \omega,  {\bm \varepsilon_{2} } )^*  \right]~, 
%
%
\label{eq:Xi}
\end{eqnarray} 
with the quantity $\Omega_{J_{i}J_{\nu}J_{f}}^{F_{i} F_{\nu}F_{f}}(  \theta, \chi, \omega,{\bm \varepsilon_{2}})$ being equal to 

\begin{eqnarray}
\Omega_{J_{i}J_{\nu}J_{f}}^{F_{i} F_{\nu}F_{f}}(  \theta, \chi, \omega,{\bm \varepsilon_{2}})&=& \sum_{L_1 \lambda_{1} L_2 \lambda_2}  (-1)^{\lambda_{2}+\lambda_{1}}[J_{\nu}, F_{\nu}]\sqrt{\left[F_{f}, F_{i},J_{f},J_{i}\right]}
\left(\begin{array}{ccc}
J_{f} & L_{2} & J_{\nu}\\
1/2 & 0 & -1/2\end{array}\right)
\left(\begin{array}{ccc}
J_{i} & L_{1} & J_{v}\\
1/2 & 0 & -1/2\end{array}\right) \nonumber \\
&&\left\{ \begin{array}{ccc}
F_{f} & L_{2} & F_{\nu}\\
J_{\nu} & I & J_{f}\end{array}\right\}
 \left\{ \begin{array}{ccc}
F_{i} & L_{1} & F_{v}\\
J_{v} & I & J_{i}\end{array}\right\}
\Theta_{L_{1}L_{2}}^{\lambda_{1},\lambda_{2}} \frac{M_{1s 2p_{3/2}}^{\lambda_{2} L_{2}}M_{1s 2p_{3/2}}^{\lambda_{1} L_{1}}}{\alpha^2 \omega^2 \mathcal{S}^2_{1s2p_{3/2}}}~.
\label{eq:omega}
\end{eqnarray}
\end{widetext}
The notation $[j1,j2,...]$ is equal to $(2j1 + 1)(2j2 + 1).$..
All dependencies of the incident polarization and geometrical settings can be traced back to the quantity $\Omega_{J_{i}J_{\nu}J_{f}}^{F_{i} F_{\nu}F_{f}}$. In particular, $\Omega_{J_{i}J_{\nu}J_{f}}^{F_{i} F_{\nu}F_{f}}$ contains the complete summation over the allowed multipoles, each one defined by $\lambda_{\gamma} L_{\gamma}$, and all multipole dependencies over $\theta$ and $\chi$. This expression follows the nomenclature as a similar nonrelativistic expression obtained in previous works \cite{abj2015, afs2015}. This complete form includes not only the multipole summation, but also the relativistic contributions that are all contained in the radial relativistic overlaps $M_{fi}^{\lambda,L}$. Explicit expressions of  $M_{1s 2p_{3/2}}^{\lambda, L}$, $\mathcal{S}_{1s2p_{3/2}}$ and $\Theta_{L_{1}L_{2}}^{\lambda_{1},\lambda_{2}}$ are given in the Appendix. There, it is also shown that $\Omega_{J_{i}J_{\nu}J_{f}}^{F_{i} F_{\nu}F_{f}}$ \eqref{eq:omega}  with both nonrelativistic and long wavelength approximations, as well as by restricting to electric dipole term is equal to Eq.~(A7) of Ref.~\cite{abj2015}.
%

%
 \begin{figure}[t]
  \centering
\includegraphics[clip=true,width=1.0\columnwidth]{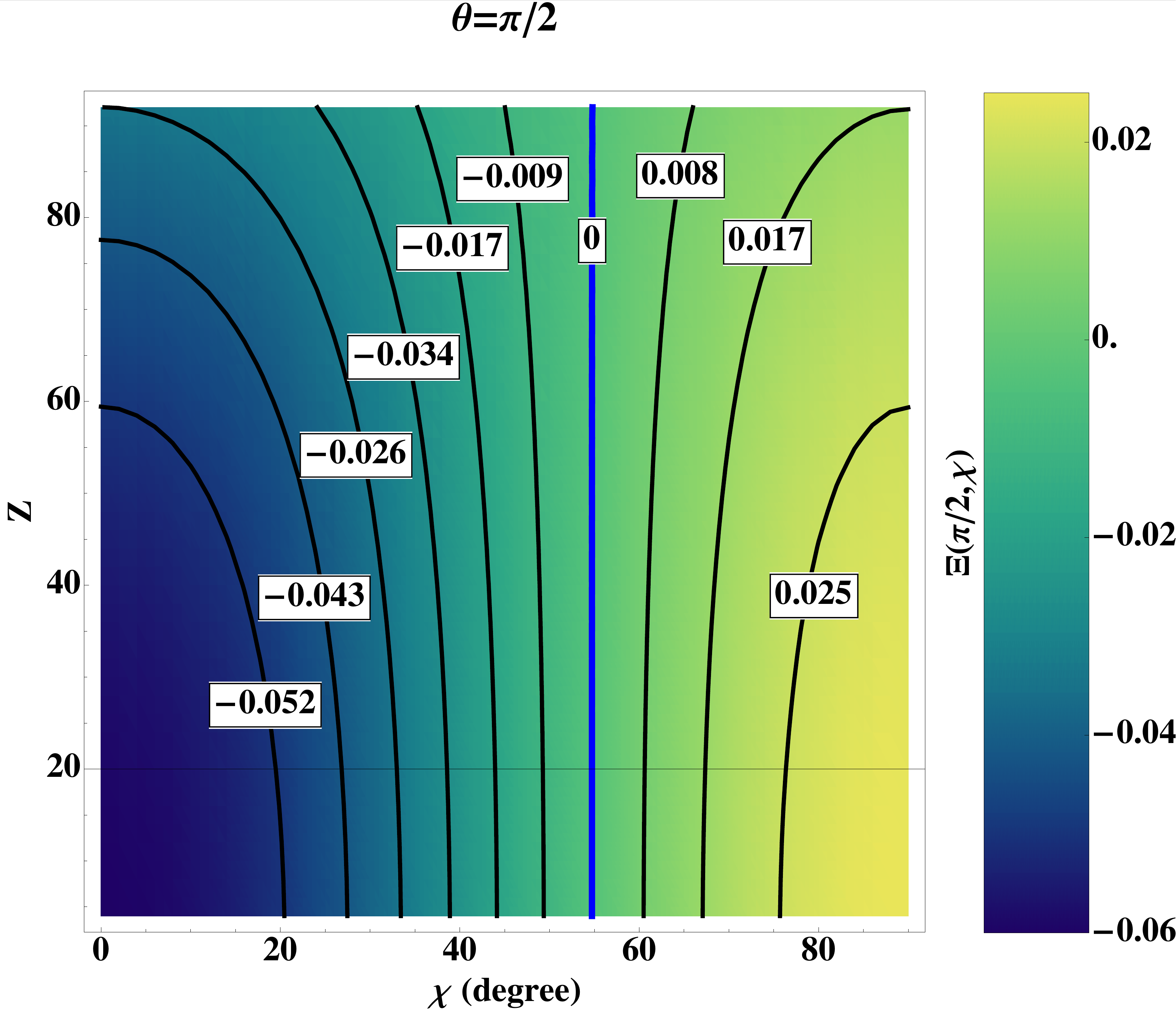}
\caption{ (Color online) Cross term $\Xi_{1/2, 3/2, 3/2}^{4,1, 2}(\pi/2, \chi)$ calculated for the lithiumlike isoelectronic sequence of ions with nuclear spin $I=1/2$.  
} 
\label{fig:z_seque}
\end{figure}

\section{Results and Discussion}
\label{sec:res_dis}

\subsection{Relativistic interference term}
\label{sec:proper_sigma}

As seen from formula \eqref{eq:Msimpli_2}, the computation of the scattering amplitudes requires knowledge about the
Lorentzian terms \eqref{eq:delta}  as well as the cross terms  \eqref{eq:Xi}.  While QI induced shifts for some lithiumlike ions will be presented and discussed later in Sec.~\ref{sec:QI_shifts}, here we will investigate the effects of relativity and nondipole terms to the cross terms \eqref{eq:Xi}, which contain the polarization and angular properties of the QI shifts. As shown in the Appendix,  the cross terms \eqref{eq:Xi}, depend only on the angular properties of the transition in a nonrelativistic limit and with only the $E1$ term. With inclusion of relativity and nondipole terms, the relativistic cross terms will have further dependencies on $\omega$. Nevertheless, $\Xi_{J_{i}J_{\nu} J_{\nu'}}^{F_{i} F_{\nu}  F_{\nu'}}$ remains independent of the relative difference between resonances and linewiths that are particular to a given atomic system. Therefore, to investigate multipole and relativistic effects to $\Xi_{J_{i}J_{\nu} J_{\nu'}}^{F_{i} F_{\nu}  F_{\nu'}}$, we can consider the isoelectronic sequence of lithiumlike isotopes, all having the same hypothetical nuclear spin, which may not be stable isotopes.  Figure~\ref{fig:z_seque} contains the values of  $\Xi_{J_{i}J_{\nu} J_{\nu'}}^{F_{i} F_{\nu}  F_{\nu'}}$ for the transition $2s\rightarrow 2p_{3/2} \rightarrow 2s$ and for this  isoelectronic sequence and having an hypothetical nuclear spin $I=1/2$ (not stable isotopes). For lower-$Z$ isotopes the dependency over $\chi$ follows the nonrelativistic limit of $\Xi_{J_{i}J_{\nu} J_{\nu'}}^{F_{i} F_{\nu}  F_{\nu'}}  \propto P_2(\cos{\chi})$ ($P_2(x)$ is the second order Legendre polynomial) \cite{bup2013}. With increasing values of $Z$, and thus, with the increase of $\omega$, the term $\Xi_{J_{i}J_{\nu} J_{\nu'}}^{F_{i} F_{\nu}  F_{\nu'}}$ deviates from the nonrelativistic case by decreasing its absolute value. The proportionality to $P_2(\cos{\chi})$ is nevertheless present even in the relativistic case of uranium lithiumlike ions. The condition of vanishing the QI shifts for $\chi'=\arccos{1/\sqrt{3}}$ thus remains valid for these atomic systems.  A close inspection of the individual multipoles shows that the biggest relativistic contribution comes from the complete form of the electric dipole term that contains retardation effects. These effects,  are only present in the radial overlaps $M_{f i}^{\lambda L}$ and thus do not influence the angular and polarization properties of the electric dipole term of $\Omega_{J_{i}J_{\nu}J_{f}}^{F_{i} F_{\nu}F_{f}}$. The next allowed multipole is a magnetic quadrupole term ($E2$) and its contribution by comparison to the allowed electric dipole is roughly 0.1\%. Therefore, no significant difference is expected with respect to the angular and polarization properties as obtained from the nonrelativistic case. 

\subsection{QI shifts in lithiumlike ions }
\label{sec:QI_shifts}

\begin{figure}[t]
  \centering
\includegraphics[clip=true,width=1.0\columnwidth]{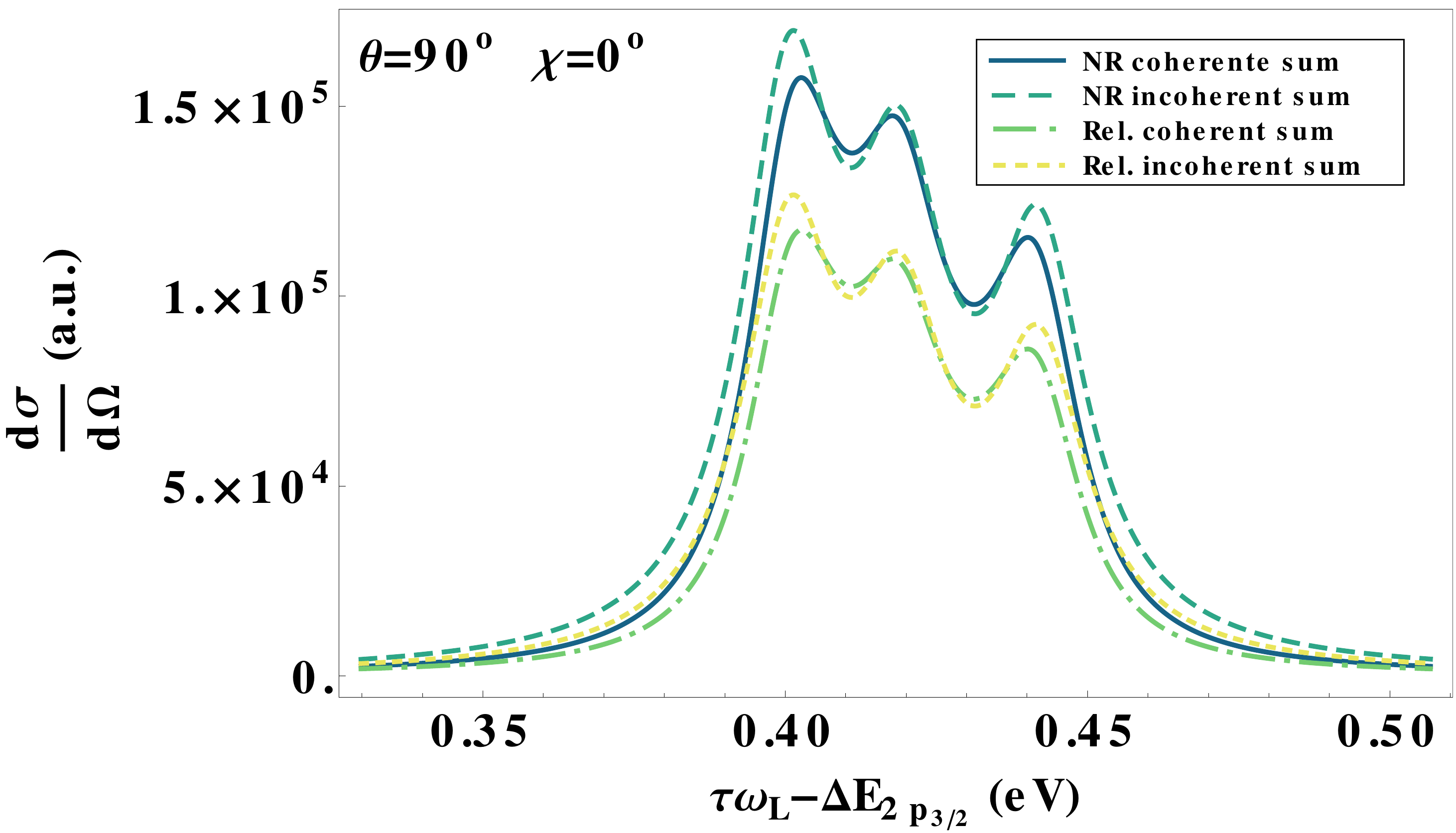}
\caption{ (Color online)  Differential cross-section \eqref{eq:Msimpli_2} for $2s_{1/2}^{F_i=4} \rightarrow 2p_{3/2} \rightarrow1s$ hyperfine resonance structure of $^{209}$Bi$^{+80}$ with $\theta=90^0$ and $\chi=90^o$.  Solid and dashed lines corresponds to the intensity with the evaluation of Eq.~\eqref{eq:Msimpli_2} with the nonrelativistic and dipole limits, while the dashed-point and point lines addresses the full relativistic and higher-multipole evaluation of Eq.~\ref{eq:Msimpli_2}. In both cases, the coherent sum corresponds to the full evaluation of Eq.~\eqref{eq:Msimpli_2}, while the incoherent sum sets the interference terms \eqref{eq:Xi} to zero. $\omega_L$ is the energy of the incident photon set to fractions of eV by a relativistic factor $\tau$.
} 
\label{fig:cross_sect}
\end{figure}

Figure \ref{fig:cross_sect} displays the differential cross section \eqref{eq:Msimpli_2} for the $2s \rightarrow 2p_{3/2} \rightarrow1s$ hyperfine resonance structure of $^{209}$Bi$^{+80}$, including the full coherent sum \eqref{eq:Msimpli_2} and having the summation restricted to just the Lorentzian terms, i.e. neglecting the cross terms. Both nonrelativistic and relativistic cases are also shown for comparison. In the same manner as $\Xi_{J_{i}J_{\nu} J_{\nu'}}^{F_{i} F_{\nu}  F_{\nu'}}$, the contribution of relativity to the cross section decreases by 35\%, as can be observed in Fig.~\ref{fig:cross_sect}. From the figure one may observe that in this set of resonances, QI's are noticeable due to the resonances having energy gaps similar to the linewidth.

\begin{table}[t]
\caption{\label{tab:shifts} Shift of the line center due to QI contribution for $^{141}$Pr$^{+56}$, $^{165}$Ho$^{+64}$, $^{207}$Pb$^{+79}$ and $^{209}$Bi$^{+80}$ for the vertical ($\delta_{\tiny \mbox{QI}}^{\perp}$) and horizontal ($\delta_{\tiny \mbox{QI}}^{\parallel}$) polarization cases ($\chi=90$\degree and $\chi=0$\degree). $i$ and $\nu$ stand for the initial and main resonant state, respectively. Values of $\delta$ relative to $\Gamma$ are given in percentage.
} 
\begin{ruledtabular}
\begin{tabular}{llllD{.}{.}{3}D{.}{.}{3}}
& I &  $i$ & $\nu$   &   \multicolumn{1}{l}{ $\bar{\delta}_{\tiny \mbox{QI}}^\parallel$ (\%) }&   \multicolumn{1}{l}{ $\bar{\delta}_{\tiny \mbox{QI}}^\perp$ (\%) } \\

$^{141}$Pr$^{+56}$ & $\frac{5}{2}$ & $2s_{1/2}^{F_i=2}$ & $2p_{3/2}^{F=1}$ & 1.3 & -0.7\\
&& & $2p_{3/2}^{F=2}$ & 0.1 & 0.0\\
& && $2p_{3/2}^{F=3}$ & -1.3 & 0.7\\
 \\[-2.0ex]  \cline{3-6} \\[-2.0ex]	\\[-2.0ex]
& &$2s_{1/2}^{F_i=3}$ &$ 2p_{3/2}^{F=2}$ & 0.3 & -0.1\\
& & &$2p_{3/2}^{F=3}$ & 1.6 & -0.7\\
& & &$2p_{3/2}^{F=4}$ & -0.8 & 0.3\\
  \\[-2.0ex]        \hline	 \\[-2.0ex] 	
$^{165}$Ho$^{+64}$ & $\frac{7}{2}$ &$2s_{1/2}^{F_i=3}$ & $2p_{3/2}^{F=2}$ & 3.1 & -1.5\\
&& & $2p_{3/2}^{F=3}$ & -1.0 & 0.3\\
&& & $2p_{3/2}^{F=4}$ & -2.9 & 1.6\\
 \\[-2.0ex]  \cline{3-6} \\[-2.0ex]	\\[-2.0ex]	
& &$2s_{1/2}^{F_i=4}$ & $2p_{3/2}^{F=3}$ & 1.0 & -0.7\\
& && $2p_{3/2}^{F=4}$ & 3.7 & -1.5\\
& && $2p_{3/2}^{F=5}$ & -2.1 & 0.8\\
  \\[-2.0ex]        \hline	 \\[-2.0ex] 	
$^{207}$Pb$^{+79}$& $\frac{1}{2}$ & $2s_{1/2}^{F_i=1}$ & $2p_{3/2}^{F=1}$ & 9.8 & -3.0\\
	 &        & & $2p_{3/2}^{F=2}$ & -2.6 & 0.9\\
	             \\[-2.0ex]        \hline	 \\[-2.0ex] 
$^{209}$Bi$^{+80}$ & $\frac{9}{2}$ & $2s_{1/2}^{F_i=4}$ & $2p_{3/2}^{F= 3}$ & 5.1 & -2.0\\
& & &$2p_{3/2}^{F= 4}$ & -3.6 & 1.0\\
& & &$2p_{3/2}^{F= 5}$ & -7.1 & 3.1\\
 \\[-2.0ex]  \cline{3-6} \\[-2.0ex]	\\[-2.0ex]	
& & $2s_{1/2}^{F_i=5}$ & $2p_{3/2}^{F= 4}$ & 2.5 & 1.0\\
& && $2p_{3/2}^{F= 5}$ & 11.1 & -3.1\\
& && $2p_{3/2}^{F= 6}$ & -6.4 & 2.5\\		        	
\end{tabular}
\end{ruledtabular}
\end{table}

  As it is observed, the influence of QI is more noticeable in regions between resonances, where no dominant excitation channels exist. Close to resonances, the influence of QI is equivalent to shifting the peak position. 
  These shifts were determined by fitting a profile generated from Eq.~\eqref{eq:Msimpli_2} with a sum of Lorentzians. Notice that if the interference terms \eqref{eq:Xi} were null, such fit would render almost zero shifts (within numerical accuracy).  The shifts normalized to $\Gamma_\nu$ are given in Table~\ref{tab:shifts} for a set of stable isotopes and for the $2p_{3/2}$ resonances. With few exceptions, all shifts are within 3\%. The cases of $^{207}$Pb$^{+79}$  $2p_{3/2}^{F=1}$ and $^{209}$Bi$^{+80}$ $2p_{3/2}^{F= 5}$ contains shifts up to 10\% of the linewidth due to the contribution of a more dominant resonance nearby.

\begin{table*}[t]
\caption{\label{tab:coe_para_del_Xi} Values of the coefficients $a_0$, $a_2$, $b_0$ and $b_2$ corresponding to the parameterizations, $\Lambda_{J_{i}J_{\nu}}^{F_{i} F_{\nu}}( \theta, \chi)=a_0 + a_2 P_2 (\cos \gamma)$ and $\Xi_{J_{i}J_{\nu} J_{\nu'}}^{F_{i} F_{\nu}  F_{\nu'}}( \theta, \chi)=b_0+b_2 P_2 (\cos \gamma)$ for $^{207}$Pb$^{79+}$ with $F_i=0$ and for $^{209}$Bi$^{80+}$ with $F_i=4$. Values are listed for both nonrelativistic (NR) and relativistic case (R). The NR values are exact and the parenthesis represents a repeating decimal. 
} 
\begin{ruledtabular}
\begin{tabular}{llllllllllll}
 &  $F_\nu$  &	 $J_\nu$	 & $F_{\nu'}$ &	 $J_{\nu'}$  & $ a_0^{\tiny\mbox{NR}}$ & $ a_2^{\tiny \mbox{NR}}$  &  $ b_2^{\tiny\mbox{NR}}$  &  $ a_0^{\tiny\mbox{R}}$  & $ a_2^{\tiny\mbox{R}}$	 & $ b_0^{\tiny\mbox{R}}$ & $ b_2^{\tiny\mbox{R}}$  \\
$^{207}$Pb$^{79+}$	&	1	&	 1/2	&	1	&	 3/2	&	 $2.7(7)\times 10^{-2}$	&	\phantom{-}0\phantom{.00$\times10^{-0}$}	&	 -$2.7(7)\times 10^{-2}$	&	 1.25$\times10^{-2}$	&	\phantom{-}0\phantom{.00$\times10^{-0}$}	&	\phantom{-}0\phantom{.00$\times10^{-0}$}	&	 -1.69$\times10^{-2}$	 \\
	&	1	&	 3/2	&	1	&	 1/2	&	 $5.5(5)\times 10^{-2}$	&	 -$2.7(7)\times 10^{-2}$	&	 -$2.7(7)\times 10^{-2}$	&	 2.40$\times10^{-2}$	&	 -2.08$\times10^{-2}$	&	\phantom{-}0\phantom{.00$\times10^{-0}$}	&	 -1.55$\times10^{-2}$	 \\
	\\[-2.0ex] \hline \\[-2.0ex]  																						
$^{209}$Bi$^{80+}$ 	&	3	 &	 3/2	 &	4	 &	 1/2	 &	1.75$\times10^{0}$	 &	 -7.291(6)$\times10^{-2}$	 &	-1.05$\times10^{-1}$	 &	 1.30$\times10^{0}$	 &	 -5.47$\times10^{-2}$	 &	 -3.61$\times10^{-5}$	 &	 -5.75$\times10^{-2}$	 \\
	&	3	 &	 3/2	 &	4	 &	 3/2	 &	\phantom{---}-	 &	\phantom{---}-	 &	-2.8875$\times10^{-1}$	 &	\phantom{---}-	 &	\phantom{---}-	 &	 -4.28$\times10^{-4}$	 &	 -2.17$\times10^{-1}$	 \\
	&	3	 &	 3/2	 &	5	 &	 1/2	 &	\phantom{---}-	 &	\phantom{---}-	 &	-7.7$\times10^{-1}$	 &	\phantom{---}-	 &	\phantom{---}-	 &	 \phantom{-}1.13$\times10^{-4}$	 &	 -4.22$\times10^{-1}$	 \\
	&	3	 &	 3/2	 &	5	 &	 3/2	 &	\phantom{---}-	 &	\phantom{---}-	 &	 -5.1(3)$\times10^{-1}$	 &	\phantom{---}-	 &	\phantom{---}-	 &	 -4.15$\times10^{-4}$	 &	 -3.85$\times10^{-1}$	 \\
	&	4	 &	 1/2	 &	3	 &	 3/2	 &	6.0$\times10^{-1}$	 &	\phantom{-}0	 &	-1.05$\times10^{-1}$	 &	 2.91$\times10^{-1}$	 &	\phantom{-}0	 &	 -2.81$\times10^{-7}$	 &	 -6.34$\times10^{-2}$	 \\
	&	4	 &	 1/2	 &	4	 &	 3/2	 &	\phantom{---}-	 &	\phantom{---}-	 &	-2.31$\times10^{-1}$	 &	\phantom{---}-	 &	\phantom{---}-	 &	 -3.61$\times10^{-7}$	 &	 -1.39$\times10^{-1}$	 \\
	&	4	 &	 1/2	 &	5	 &	 3/2	 &	\phantom{---}-	 &	\phantom{---}-	 &	-2.64$\times10^{-1}$	 &	\phantom{---}-	 &	\phantom{---}-	 &	\phantom{-}5.05$\times10^{-7}$ 	 &	 -1.59$\times10^{-1}$	 \\
	&	4	 &	 3/2	 &	3	 &	 3/2	 &	1.65$\times10^{0}$	 &	-4.0425$\times10^{-1}$	 &	-0.28875	 &	 1.23$\times10^{0}$	 &	 -3.04$\times10^{-1}$	 &	 -4.28$\times10^{-4}$	 &	 -2.17$\times10^{-1}$	 \\
	&	4	 &	 3/2	 &	4	 &	 1/2	 &	\phantom{---}-	 &	\phantom{---}-	 &	-2.31$\times10^{-1}$	 &	\phantom{---}-	 &	\phantom{---}-	 &	 -4.64$\times10^{-5}$	 &	 -1.26$\times10^{-1}$	 \\
	&	4	 &	 3/2	 &	5	 &	 1/2	 &	\phantom{---}-	 &	\phantom{---}-	 &	-5.94$\times10^{-1}$	 &	\phantom{---}-	 &	\phantom{---}-	 &	\phantom{-}2.79$\times10^{-4}$	 &	 -3.26$\times10^{-1}$	 \\
	&	4	 &	 3/2	 &	5	 &	 3/2	 &	\phantom{---}-	 &	\phantom{---}-	 &	-1.32$\times10^{-1}$	 &	\phantom{---}-	 &	\phantom{---}-	 &	\phantom{-}1.29$\times10^{-4}$	 &	 -9.93$\times10^{-2}$	 \\
	&	5	 &	 1/2	 &	3	 &	 3/2	 &	1.65$\times10^{0}$	 &	\phantom{-}0	 &	-7.7$\times10^{-1}$	 &	 7.99$\times10^{-1}$	 &	\phantom{-}0	 &	\phantom{-}8.86$\times10^{-7}$	 &	 -4.65$\times10^{-1}$	 \\
	&	5	 &	 1/2	 &	4	 &	 3/2	 &	\phantom{---}-	 &	\phantom{---}-	 &	-5.94$\times10^{-1}$	 &	\phantom{---}-	 &	\phantom{---}-	 &	\phantom{-}2.18$\times10^{-6}$	 &	 -3.59$\times10^{-1}$	 \\
	&	5	 &	 1/2	 &	5	 &	 3/2	 &	\phantom{---}-	 &	\phantom{---}-	 &	-2.86$\times10^{-1}$	 &	\phantom{---}-	 &	\phantom{---}-	 &	\phantom{-}3.29$\times10^{-6}$	 &	 -1.73$\times10^{-1}$	 \\
	&	5	 &	 3/2	 &	3	 &	 3/2	 &	1.1$\times10^{0}$	 &	\phantom{-}9.5(3)$\times10^{-2}$	 &	-	 &	 8.20$\times10^{-1}$	 &	 \phantom{-}7.20$\times10^{-2}$	 &	\phantom{---}-	 &	\phantom{---}-	 \\
	&	5	 &	 3/2	 &	4	 &	 1/2	 &	\phantom{---}-	 &	\phantom{---}-	 &	-2.64$\times10^{-1}$	 &	\phantom{---}-	 &	\phantom{---}-	 &	\phantom{-}6.48$\times10^{-5}$	 &	 -1.45$\times10^{-1}$	 \\
	&	5	 &	 3/2	 &	4	 &	 3/2	 &	\phantom{---}-	 &	\phantom{---}-	 &	-	 &	\phantom{---}-	 &	\phantom{---}-	 &	\phantom{---}-	 &	\phantom{---}-	 \\
	&	5	 &	 3/2	 &	5	 &	 1/2	 &	\phantom{---}-	 &	\phantom{---}-	 &	-2.86$\times10^{-1}$	 &	\phantom{---}-	 &	\phantom{---}-	 &	\phantom{-}4.22$\times10^{-4}$	 &	 -1.57$\times10^{-1}$	 \\
\end{tabular}
\end{ruledtabular}
\end{table*}

\section{Conclusion}  
\label{sec:sum}

We performed a theoretical investigation of the quantum interference for the hyperfine structure of highly charged isotopes, measurable by laser spectroscopy. Full relativistic and multipole frameworks are employed here. While multipole contribution was quantified to be negligible, retardation effects to the electric dipole term contributes to a reduction of the cross section by 35\%. Moreover, we verify that the polarization and  geometrical dependencies of quantum interference remains unaffected by inclusion of relativity and nondipole terms.

We quantified the effect of quantum interference on the line center of several resonances of a few selected highly charged isotopes and observed systematic shifts up to 10\% if such resonances were to be fitted by regular Lorentzian profiles.

\appendix* 
\section{Hyperfine multipole matrix elements}
\label{sec:append}


One photon multipole matrix elements have been evaluated several times in the literature \cite{gra1974, sfi1998, asf2009}, therefore we give only a brief description of the expressions employed in this work. The relativistic multipole matrix elements $\langle f |\bm{ \alpha\cdot a_{LM}^{(\lambda)*}(\hat{r})}| i \rangle$, which are obtained after the inclusion of the spherical multipole expansion \eqref{eq:multexpanc} in Eq.~\eqref{fig:transit}, can be simplified with the use of the Wigner-Eckart theorem \cite{ros1957} and by considering the overall atomic state being the product coupling of the nucleus and electron angular momenta, given by
\begin{equation}
\left|\beta F m\right\rangle =\sum_{m_{I}m_{J}}\left\langle J m_{J} I m_{I} | F m \right\rangle \left|I m_{I} \right\rangle \left|\beta J m_{J}\right\rangle ~.
\end{equation}
The quantities  $\left\langle j_1 m_{j_1} j_2 m_{j_2} | j_3 m_{j_3} \right\rangle$ are standard Clebsch-Gordan coefficients.
After additional angular momentum algebra \cite{ros1957}, the atomic multipole matrix elements are explicitly given by (velocity gauge)
%
\begin{eqnarray}
&&\langle f |\bm{ \alpha\cdot a_{LM}^{(\lambda)*}(\hat{r})}| i \rangle= \\
&&(-1)^{F_{f}+I+F_{i}+2J_{f}+\lambda-m_{f}-1/2}
\sqrt{\frac{4 \pi \left[F_{f},F_{i},J_{f},J_{i}\right]}{[L]}}  \times \nonumber \\
&&\left(\begin{array}{ccc}
F_{f} & L & F_{i}\\
-m_{f} & M & m_{i}\end{array}\right)
\left\{ \begin{array}{ccc}
J_{i} & I & F_{i}\\
F_{f} & L & J_{f}\end{array}\right\}  \times \nonumber \\
&& \left(\begin{array}{ccc}
J_{f} & L & J_{i}\\
1/2 & 0 & -1/2\end{array}\right)M_{fi}^{\lambda L} ~.
\label{eq:onematrix}
\end{eqnarray}
%
%
where the notation $[j_1, j_2,...]$ is equal to $(2j_1+1)(2j_2+1)...~$. The relativistic radial overlaps $M_{fi}^{\lambda L}$  ($\lambda=0,1$) have the following expressions,

\begin{eqnarray}
M_{fi}^{1L} &=&\left( \frac{L}{L+1}\right) ^{1/2}\left[
\left( \kappa_{f}-\kappa_{i}\right) I_{L+1}^{+}+\left( L+1\right) I_{L+1}^{-}%
\right]  \nonumber \\
&&-\left( \frac{L+1}{L}\right) ^{1/2}\left[ \left( \kappa_{f}-\kappa_{i}\right)
I_{L-1}^{+}-LI_{L-1}^{-}\right]~, \nonumber \\
\label{M_al_bet1}
\end{eqnarray}
and 
\begin{equation}
M_{fi}^{0L}=\frac{2L+1}{\left[ L\left( L+1\right) \right]
^{1/2}}\left( \kappa_{f}+\kappa_{i}\right) I_{L}^{+} , 
\label{M_al_bet}
\end{equation}
for both electric ($\lambda=1$) and magnetic type contributions ($\lambda=1$), respectively, with $I_{L}^\pm=\int_{0}^{\infty }\left( P_{f}Q_{i}\pm P_{i}Q_{f}\right)
j_{L}\left( \frac{\omega r}{c}\right) dr$. Here, $P$ and $Q$ stands for the large and small components of
the radial Dirac wave function with relativistic angular momentum $\kappa$, and $j_L(x)$ is the spherical Bessel function of first kind. In case of long wavelength approximation ($\omega r/c  \simeq 0$)  this function approximates to $j_L(\omega r/c)\simeq (\omega r/c)^L/[L]!!$, which  simplifies the electric dipole radial overlap between states $2p_{3/2}$ and $1s$ to $M_{1s2p_{3/2}}^{11}\simeq -\sqrt{2} (I_0^+-I_0^-)$. Moreover, by only considering the term of order $1/c$ of $Q$ and the commutator $\left[ \frac{d^2}{2dr^2}, r\right]=\frac{d}{dr}$, the quantity $M_{1s2p_{3/2}}^{11}$ further simplifies to 
\begin{eqnarray}
M_{1s2p_{3/2}}^{11}\simeq -\alpha \omega \sqrt{2} \int_0^{\infty} P_{1s}P_{2p_{3/2}}rdr =  -\alpha \omega \sqrt{2} \mathcal{S}_{\nu i}~. \nonumber \\ 
\label{eq:M_E1simple}
\end{eqnarray}
After inserting the matrix elements \eqref{eq:onematrix} into Eq.~\eqref{eq:Msimpli} and rearranging the terms we get Eq.~\eqref{eq:omega}, where the remained angular term $\Theta_{L_{1}L_{2}}^{\lambda_{1},\lambda_{2}}$ is given by
\begin{eqnarray}
&&\Theta_{L_{1}L_{2}}^{\lambda_{1},\lambda_{2}}= \frac{4\pi }{\sqrt{\left[L_{2},L_{1}\right]}} \sum_{M_{1}M_{2}}\sum_{m_{\nu}}  (-1)^{m_\nu + m_f +1} \times \nonumber \\
&&
\left(\begin{array}{ccc}
F_{f} & L_{2} & F_{v}\\
-m_{f} & M_{2} & m_\nu
\end{array}\right) 
\left(\begin{array}{ccc}
F_{i} & L_{1} & F_{v}\\
-m_{i} & M_{1} & m_\nu \end{array}\right)\times \nonumber \\
&&\left[\bm{\varepsilon_{2}\cdot}\bm{Y}_{L_{2}M_{2}}^{(\lambda_{2})}(\hat{\bm{k}_{2}})\right]\left[\bm{\varepsilon_{1}\cdot}\bm{Y}_{L_{1}M_{1}}^{*(\lambda_{1})}(\hat{\bm{k}_{1}})\right]~.
\end{eqnarray}
We use here the standard normalization of the spherical harmonics ($Y_{00}(\theta,\phi)=\frac{1}{2\sqrt{\pi}}$), therefore for the electric dipole term ($L=1$, $\lambda=1$) we found  that
\begin{equation}
\bm{\varepsilon}\cdot\bm{Y}_{1M}^{(1)*}=(-1)^{M}\frac{1}{\sqrt{2}}\sqrt{\frac{3}{4\pi}} \varepsilon^{(-M)}~,
\label{eq:angular_E1sim}
\end{equation}
where $\varepsilon^{(-M)}$ are the spherical form of the (normalized) polarization vector. By considering  previous result \eqref{eq:angular_E1sim} the quantity $\Theta_{L_{1}L_{2}}^{\lambda_{1},\lambda_{2}}$ is half of Eq. (A8) in \cite{abj2015}. With Eqs.~\eqref{eq:M_E1simple} and \eqref{eq:angular_E1sim}, the quantity $\Omega_{J_{i}J_{\nu}J_{f}}^{F_{i} F_{\nu}F_{f}}(  \theta, \chi,{\bm \varepsilon_{2}})$ approximates to  Eq. (A7) of \cite{abj2015}. 



%
\begin{acknowledgments}

This work was funded by the Portuguese Funda\c{c}\~{a}o para a Ci\^{e}ncia e a Tecnologia (FCT/MCTES/PIDDAC) under grant UID/FIS/04559/2013
(LIBPhys).
P.~A. acknowledges the support of the FCT, under Contracts No. SFRH/BPD/92329/2013.  
L. S. acknowledges financial support from the People Programme (Marie Curie Actions) of the European Union's Seventh Framework Programme (\emph{FP7/2007-2013}) under REA Grant Agreement No. [\emph{291734}]. 
Laboratoire Kastler Brossel (LKB) is ``Unit\'e Mixte de Recherche de Sorbonne Universit\'e, de ENS-PSL Research University, du Coll\`{e}ge de France et du CNRS n$^{\circ}$ 8552''.
\end{acknowledgments}

\bibliography{QIR_articles}


\end{document}